\documentclass[journal]{IEEEtran}

\ifCLASSINFOpdf
\else
   \usepackage[dvips]{graphicx}
\fi
\usepackage{url}

\usepackage{graphicx}
\usepackage{amsmath,graphicx}
\usepackage{amssymb}
\usepackage{algorithm}
\usepackage{algorithmic}
\usepackage{setspace}
\usepackage{stfloats}
\usepackage{subfigure}
\usepackage{balance}
\begin{document}

\title{Constellation Design for Deep Joint Source-Channel Coding}

\author{Mengyang Wang, Jiahui Li, \IEEEmembership{Member, IEEE}, Mengyao Ma, \IEEEmembership{Member, IEEE}, \\and Xiaopeng Fan, \IEEEmembership{Senior Member, IEEE}\vspace{-2mm}
\thanks{This work was supported in part by the National Science Foundation of China (NSFC) under grants 61972115 and 61872116. \textit{(Corresponding author: Xiaopeng Fan.)}}
\thanks{Mengyang Wang is with the School of Computer Science, Harbin Institute of Technology, Harbin 150001, China, and also with the Wireless Technology Lab, Huawei, Shenzhen 518129, China. Jiahui Li and Mengyao Ma are with the Wireless Technology Lab, Huawei, Shenzhen 518129, China (e-mail: lijiahui666@huawei.com; ma.mengyao@huawei.com). Xiaopeng Fan is with the School of Computer Science, Harbin Institute of Technology, Harbin 150001, China (e-mail: fxp@hit.edu.cn).}
}

\maketitle

\begin{abstract}
Deep learning-based joint source-channel coding (JSCC) has shown excellent performance in image and feature transmission. However, the output values of the JSCC encoder are continuous, which makes the constellation of modulation complex and dense. It is hard and expensive to design radio frequency chains for transmitting such full-resolution constellation points. In this paper, two methods of mapping the full-resolution constellation to finite constellation are proposed for real system implementation. The constellation mapping results of the proposed methods correspond to regular constellation and irregular constellation, respectively. We apply the methods to existing deep JSCC models and evaluate them on AWGN channels with different signal-to-noise ratios (SNRs). Experimental results show that the proposed methods outperform the traditional uniform quadrature amplitude modulation (QAM) constellation mapping method by only adding a few additional parameters.
\end{abstract}


\IEEEpeerreviewmaketitle

\section{Introduction}
\label{sec:1}
In traditional communication systems, source coding and channel coding are two separate steps. Source coding is used to remove source redundancy, while channel coding is used to add redundant information to enhance robustness. Shannon’s separation theorem proves that this two-step approach  is optimal theoretically in the asymptotic limit of infinitely long source and channel
blocks \cite{1}. However, previous research proves that joint source-channel coding (JSCC) outperforms the separate source coding and channel coding in the finite blocklength reigme \cite{2}. 

Nowadays, with the development of deep learning (DL) and convolutional neural networks (CNNs) \cite{3}, some CNN-based JSCC models are proposed to transmit images and deep intermediate features. In \cite{4} and \cite{5}, CNN-based autoencoders (AEs) were used as the transmitter and receiver for the images. Images are encoded and transmitted at the transmitter, and the transmitted signals are received and reconstructed at the receiver after passing through noisy channels. Based on \cite{4} and \cite{5}, \cite{6} added a feedback model to improve the performance and robustness. Considering the change of SNR, \cite{7} proposed a SNR-adaptive image transmission model. By training the model with channels under different SNRs, good performance is obtained. Moreover, inspired by collaborative intelligence (CI) \cite{17}, where the deep model can be split into the mobile device and the edge server to balance the computational load, \cite{8} proposed BottleNet++ to compress and transmit the intermediate feature of the ResNet \cite{9} over noisy channels. To improve the computing capability of the receiver, \cite{10} proposed an asymmetrical AE-based JSCC model for transmitting the deep features of the VGG network \cite{11}. The AE-based JSCC model was also applied to person re-identification (re-ID) \cite{12}. All of these models are single-task learning (STL) models, and \cite{13} proposed a multi-task learning (MTL) JSCC model that maintains good performance on both tasks.

Although these JSCC models achieve excellent results, few models consider constellation design issues in the modulation module of communication systems. The output values of the JSCC encoder are continuous, so the full-resolution constellation is complex and dense, which is hard and expensive to implement in radio frequency (RF) systems. Moreover, transmitting the continuous values is also a big burden for the antenna. To quantize the full-resolution constellation into a constellation with finite points, there are some constellation design approaches. In \cite{14}, deep learning was applied to the physical layer to transmit the messages, where a fully connected neural network (FCNN) was used to build the transmitter and receiver. There are two output nodes in the last layer of the transmitter, which corresponds to the real and imaginary parts of the constellation. Inspired by \cite{14}, a higher-order constellation design model was implemented by FCNN in \cite{15}. Apart from the constellation design of the physical layer, \cite{16} proposed a semantic communication system, and the constellation of coded information is mapped to a finite set by uniform quantization.

To achieve more robust and effective constellation mapping of the full-resolution constellation, two approaches are proposed in this paper. By adding a few learnable parameters to the deep JSCC model and fine-tuning the model, both regular and irregular finite constellations are obtained. To the best of the authors' knowledge, this is the first work to perform learning-based constellation design for deep JSCC. Our main contributions are summarized as follows:
\begin{enumerate}
\item We propose a deep JSCC model with constellation mapping, which is suitable for real system implementation.
\item Two different constellation mapping methods are proposed by adding a few learnable parameters, which do not increase the complexity much.
\item The two constellation mapping methods are evaluated, and their performance and robustness outperform the traditional uniform quadrature amplitude modulation (QAM) constellation mapping.
\end{enumerate}

The rest of the paper is organized as follows. In Section \ref{sec:2}, we introduce the system background, including the whole network architecture, constellation visualization and QAM mapping. The proposed constellation mapping methods are introduced in Section \ref{sec:4}. Experimental results are discussed in Section \ref{sec:5}. Finally, the paper is concluded in Section \ref{sec:6}.

\emph{Notations:} In constellation mapping, each constellation point \textsl{p} is a complex number, Re\{\textsl{p}\} and Img\{\textsl{p}\} denote the real and imaginary part of \textsl{p}, respectively. $\textsl{F} \subset \mathbb{C}$ denotes the set of full-resolution constellation points, and $\textsl{C} \subset \mathbb{C}$ denotes the set of finite constellation points. Re\{\textsl{F}\} and Img\{\textsl{F}\} $\subset \mathbb R$ denote the set of real and imaginary part of constellation points in \textsl{F}, respectively.

\section{Background}
\label{sec:2}
\subsection{Overall System Architecture}
\label{sec:2.1}
\begin{figure}[t]
\begin{minipage}[b]{1.0\linewidth}
  \centering
  \label{fig:1(a)}
  \centerline{\includegraphics[width=\linewidth]{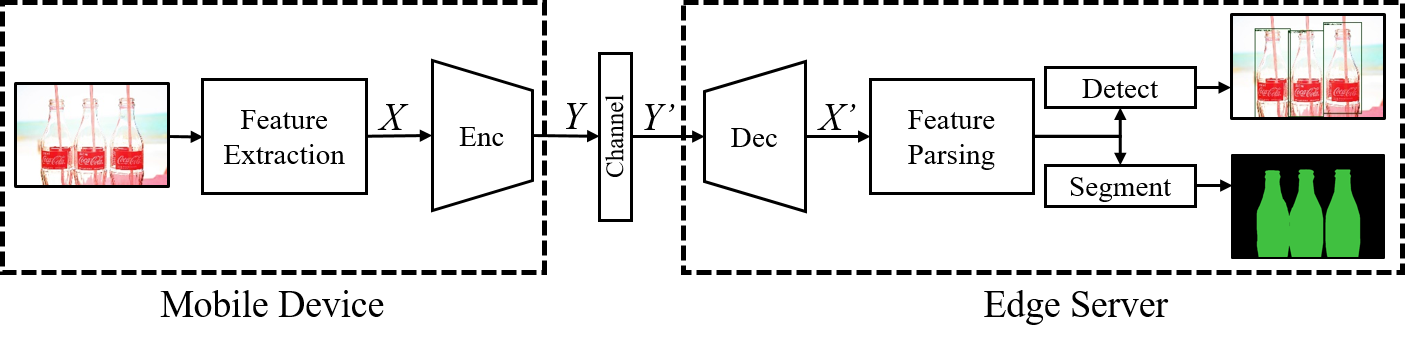}}
  \centerline{(a)}\medskip
\end{minipage}

\begin{minipage}[b]{1.0\linewidth}
  \centering
  \label{fig:1(b)}
  \centerline{\includegraphics[width=0.75\linewidth]{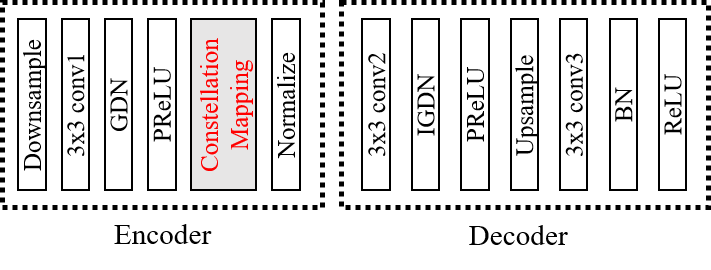}}
  \centerline{(b)}\medskip
\end{minipage}
\caption{(a): The overall architecture of MTL JSCC network. (b): The deep JSCC model.}
\label{fig:1}
\end{figure}

The overall system architecture used in this paper is shown in Fig. \ref{fig:1}, which comes from our previous work \cite{13}. The whole system consists of an MTL network and an asymmetric AE-based JSCC module. The MTL network performs object detection and semantic segmentation. The JSCC model compresses and transmits the intermediate feature of the MTL network. The wireless channel between the mobile device and the edge server is modeled by an additive white Gaussian noise (AWGN) model. Given input feature $z \in \mathbb R^{B}$ and output feature $z^{\prime} \in \mathbb R^{B}$, the transfer function of AWGN channel is written as $z^{\prime}=z+n$, with $n \sim \mathcal{N}(0, \sigma^2)$. The parameter $\sigma^2$ is the noise variance, which denotes the channel condition. Besides, to meet the average transmit power constraint of $P=1$, i.e. $\frac{1}{B} \sum\limits_{i=1}^{B} z_i^2 = P$, a power normalization layer is put at the end of the encoder. The MTL JSCC model in \cite{13} can support 512$\times$ compression for the intermediate feature when trained under AWGN channel. We use the well-trained model to illustrate the proposed constellation mapping methods and verify their effectiveness.

The constellation mapping step is added before the normalization layer in the encoder, which maps the full-resolution constellation to the finite constellation while satisfying power constraint. The constellation mapping methods can also be used in other scenarios, such as STL JSCC.

\subsection{Full-resolution constellation and QAM mapping}
\label{sec:2.2}
\begin{figure}[t]
\subfigure{
\begin{minipage}[b]{0.46\linewidth}
  \centering
  \centerline{\includegraphics[width=0.95\linewidth]{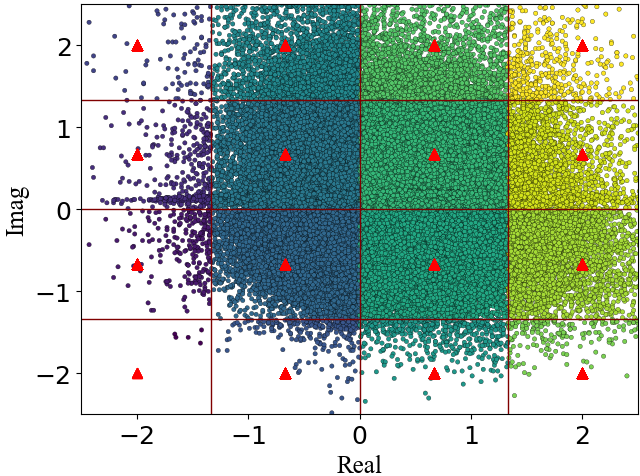}}
  \centerline{(a)}\medskip
\end{minipage}
\label{fig:2(a)}
}
\hfill
\subfigure{
\begin{minipage}[b]{0.46\linewidth}
  \centering
  \centerline{\includegraphics[width=0.95\linewidth]{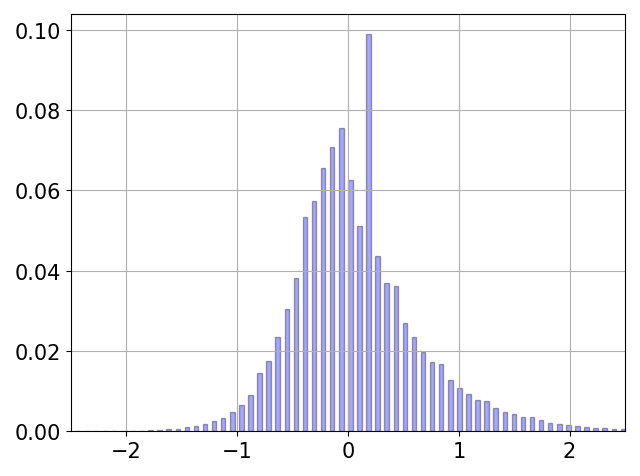}}
  \centerline{(b)}\medskip
\end{minipage}
\label{fig:2(b)}
}
\caption{(a): Full-resolution constellation and 16QAM mapping. \\(b): Distribution of JSCC encoder output.}
\label{fig:2}
\end{figure}

We visualize the full-resolution constellation of the JSCC encoder outputs by the visualization method in \cite{14}. Suppose the original dimension of the JSCC encoder outputs is (H, W, C), we reshape it to (H*W*C/2, 2). The last dimension corresponds to the real and imaginary parts of the constellation. From Fig. \ref{fig:2(a)}, we observe that the points in the full-resolution constellation are very dense, making it difficult to design suitable antennas and RF chains in practice. To facilitate the implementation of the deep JSCC model, it is necessary to quantize the full-resolution constellation into a finite constellation. Furthermore, from the distribution of JSCC encoder outputs in Fig. \ref{fig:2(b)}, we find that the values are mainly distributed in [-2, 2]. Therefore, JSCC encoder outputs are clipped into [-2, 2] before quantization to prevent outliers from affecting the final performance.

The traditional constellation mapping method is QAM mapping, which applies uniform quantization to the real and imaginary parts of the full-resolution constellation. The constellation points of QAM are finite and regularly distributed. 16QAM mapping is also shown in Fig. \ref{fig:2(a)}, which applies 2-bits uniform quantization to the real and imaginary parts of the full-resolution constellation. 16QAM divides the full-resolution constellation into 16 rectangular clusters, with points in each cluster mapped to a finite constellation point represented by a red triangle. 

However, points in full-resolution constellations may not be evenly distributed in most deep JSCC models as shown in Fig. \ref{fig:2(b)}, so QAM mapping may not be the optimal quantization approach to maintain the high performance of tasks.

\begin{figure}[t]
\begin{minipage}{1.0\linewidth}
  \centering
  \centerline{\includegraphics[width=0.9\linewidth]{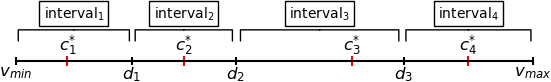}}
\end{minipage}
\caption{A simple diagram of MRC}
\label{fig:3}
\end{figure}

\section{Methodlogy}
\label{sec:4}
In this section, full-resolution constellation is divided into clusters, and each cluster contains a set of constellation points. Quantizing the full-resolution constellation to the finite constellation is treated as mapping the full-resolution constellation points in the same cluster to one finite constellation point. We propose two methods in Section \ref{sec:4.1} and \ref{sec:4.2}, respectively, to map the full-resolution constellation to regular and irregular finite constellations.

\subsection{Mapping to Regular Constellation (MRC)}
\label{sec:4.1}
In this section, we propose a method to map \textsl{F} to \textsl{C}, by mapping the data in Re\{\textsl{F}\} and Img\{\textsl{F}\} to Re\{\textsl{C}\} and Img\{\textsl{C}\}, respectively. Real part of the data, Re\{\textsl{F}\}, is taken as the example to illustrate the approach.

Suppose the range of Re\{\textsl{F}\} $\subset \mathbb{R}$ is $[v_\text{min}, v_\text{max}]$, and the mapping result set is Re\{\textsl{C}\} $=\{c_1^*, c_2^*, ..., c_M^*\} \subset \mathbb R $. We define set $D= \{d_1, d_2, ..., d_{M-1}\} \subset \mathbb R $ to divide the range \\ $[v_\text{min}, v_\text{max}]$ into $M$ intervals (i.e., $[v_\text{min}, d_1], [d_1, d_2],  ..., [d_{M-1},$\\$ v_\text{max}]$), and the input data in the $i^{th}$ interval is mapped to $c_i^*$ for $i=1,2,...,M.$ A simple diagram with $M=4$ is shown in Fig. \ref{fig:3}. The length of each interval indicates the size of the cluster in the real axis direction. In this method, we consider regular constellation points, so Re\{\textsl{C}\} is set to the same value as the uniform quantization result of Re\{\textsl{F}\} (i.e.,$ c^*_i = v_\text{min} + (i-1)*(v_\text{max}-v_\text{min})/(M-1)), i = 1, 2, \ldots, M$), while $D$ is set to be learnable to adjust the length of the interval. 

Similarly, we apply this method on Img\{\textsl{F}\} to adjust the sizes of the clusters in the imaginary axis direction. \textsl{C} can be obtained by the Cartesian product of Re\{\textsl{C}\} and Img\{\textsl{C}\}.

\begin{algorithm}
	\setstretch{0.9}
	\renewcommand{\algorithmicrequire}{\textbf{Input:}}
	\renewcommand{\algorithmicensure}{\textbf{Output:}}
	\caption{Mapping to Regular Constellation(MRC)}
	\label{alg1}
	\begin{algorithmic}[1]
         \REQUIRE $x$, Re\{\textsl{C}\}, $D$, gap$(c_k^*, c_{k+1}^*)$
         \STATE $x$ $\leftarrow$ clip( $x$, $v_\text{max} = 2, v_\text{min} = -2$ )
         \STATE \textbf{Forward Pass:}
         \STATE $k \leftarrow \mathop{\arg\min}\limits_{j}|x - d_{j}|$
         \STATE $d_{k} \leftarrow D_{[k]},c_{k}^* \leftarrow Re\{\textsl{C}\}_{[k]}$
         \STATE Result$_{forward}$ $ \leftarrow c_{k}^* + $ Heaviside($x - d_{k}$)$*$gap$(c_k^*, c_{k+1}^*)$
         \STATE \textbf{Backpropagation:}
         \STATE $\hat{d_{k}} \leftarrow \sum_{j=1}^{M-1} \frac{exp(-\delta|x-d_{j}|)}{\sum_{l=1}^{M-1}exp(-\delta|x-d_{l}|)}d_{j}$
         \STATE $\hat{c_{k}^*} \leftarrow \sum_{j=1}^{M-1} \frac{exp(-\delta|x-d_{j}|)}{\sum_{l=1}^{M-1}exp(-\delta|x-d_{l}|)}c_{j}^*$
         \STATE Result$_{backward}$ $\leftarrow \hat{c_{k}^*}$ + sigmoid($\delta(x-\hat{d_{k}})$)$*$gap$(c_k^*, c_{k+1}^*)$
		\ENSURE $tf.stopgradient$(Result$_{forward}$ - Result$_{backward}$) + Result$_{backward}$
	\end{algorithmic}  
\end{algorithm}

The pseudocode is shown in Algorithm 1, which we implement in Tensorflow \cite{19}. The working principle of forward pass is as follows: For an input value $x \in$ Re\{\textsl{F}\}, the closest $d_k$ is found, and then two adjacent intervals are determined by $d_k$. If $x>d_{k}$, which means $x \in$ interval$_{k+1}$, the mapping result is $c_{k+1}^*$, otherwise the result is $c_{k}^*$. 

Since the mapping process is non-differentiable, the soft approximation method is proposed in the backpropagation, which is inspired by\cite{18}. The weighted sums of the elements in $D$ and Re\{\textsl{C}\} are used as the soft assignment of $d_k$ and $c_k^*$, respectively, and the weights are derived from softmax based on the distances between inputs and the elements in $D$. Step function $Heaviside$ \footnote{Heaviside function value is 1 for positive arguments and 0 for others.} determines whether result is $c_k^*$ or $c_{k+1}^*$, and the `gap$(c_k^*, c_{k+1}^*)$' is the distance between $c_k^*$ and $c_{k+1}^*$. $sigmoid$ is used to perform the soft approximation of $Heaviside$. $\delta$ is a big positive number to make the Result$_{backward}$ close to Result$_{forward}$. By using $tf.stopgradient$, the forward pass output is Result$_{forward}$, and the backpropagation output is Result$_{backward}$.

Since Re\{\textsl{C}\} and Img\{\textsl{C}\} are set to the same value as the uniform quantization results of Re\{\textsl{F}\} and Img\{\textsl{F}\}, respectively, \textsl{C} is distributed as regularly as QAM. However, as MRC adjusts the sizes of the clusters in the real and imaginary axis direction, the cluster size obtained by MRC is different from that of QAM. In addition, MRC only introduces a learnable set $D$, which has little effect on the complexity of the model.

\subsection{Mapping to Irregular Constellation (MIC)}
\label{sec:4.2}
In this section, we propose a method to map \textsl{F} to \textsl{C} based on clustering, and the finite constellation points in \textsl{C} are adjustable and can be trained end-to-end. The clusters are obtained by clustering the full-resolution constellation points in \textsl{F} based on the distance to the points in \textsl{C}. 

Suppose the learnable finite constellation point set is \textsl{C} $ = \{c_1, c_2, ..., c_N\} \subset \mathbb C$, and the initial value of \textsl{C} is set to the same value as QAM. The value of \textsl{C} will be adjusted based on task performance. For an input constellation point $\textsl{p} \in \textsl{F}$, the clustering algorithm based on the Euclidean distance is used to map \textsl{p} to a nearest constellation point in \textsl{C}. As the process of finding the nearest constellation point is non-differentiable, we implement soft assignment in the backpropagation based on the method in Algorithm 1. The pseudocode is shown in Algorithm 2.

\begin{algorithm}
	\setstretch{0.9}
	\renewcommand{\algorithmicrequire}{\textbf{Input:}}
	\renewcommand{\algorithmicensure}{\textbf{Output:}}
	\caption{Mapping to Irregular Constellation(MIC)}
	\label{alg2}
	\begin{algorithmic}[1]
         \REQUIRE \textsl{p}, \textsl{C}
         \STATE \textsl{p} $\leftarrow$ clip( \textsl{p}, $v_\text{max} = 2, v_\text{min} = -2$ )
         \STATE \textbf{Forward Pass:}
         \STATE $index \leftarrow \mathop{\arg\min}\limits_{j}\sqrt{(\textsl{p} - c_{j})^2}$
         \STATE Result$_{forward}$ $ \leftarrow C_{[index]}$
         \STATE \textbf{Backpropagation:}
         \STATE Result$_{backward}$ $\leftarrow \sum_{j=1}^{N} \frac{exp(-\delta \sqrt{(\textsl{p} - c_{j})^2})}{\sum_{l=1}^{N}exp(-\delta \sqrt{(\textsl{p} - c_{l})^2})}c_{j}$
		\ENSURE $tf.stopgradient$(Result$_{forward}$ - Result$_{backward}$) + Result$_{backward}$
	\end{algorithmic}  
\end{algorithm}

$\delta$ is the same as MRC. \textsl{C} and clusters are obtained by training the model. As the constellation points in \textsl{F} are not evenly distributed, better performance can be obtained using irregularly distributed \textsl{C} and irregularly shaped clusters. In addition, MIC only adds a learnable parameter set \textsl{C}, which has little effect on the model complexity.

\section{EXPERIMENTS}
\label{sec:5}

\subsection{Dataset and training strategy}
\label{sec:5.1}

In the experiments, we use the same dataset as \cite{13}, which is part of the Open Image Dataset \cite{20}, to train the JSCC model with the addition of the constellation mapping step. Based on the trained JSCC model, the training strategy of the new model has two steps. The first step is to train the parameters in the constellation mapping part and fix other parts in Fig. \ref{fig:1}. The second step is to train the parameters of the decoder and feature parsing part. The reason why we fix the parameters of the feature extraction part and the encoder part is that the output data distribution of JSCC encoder should be constant.

The loss function to train the new model is the same as the JSCC model, where the loss function $\mathcal L_{MTL}$ has two parts, i.e., the loss function of object detection $\mathcal L_{det}$ and semantic segmentation $\mathcal L_{seg}$:
\begin{equation}
\mathcal L_{MTL} = \mathcal L_{det} + \mathcal L_{seg}.
\end{equation}
The model is optimized by the Adam algorithm with a min-batch size of 32. The iteration number of the first training step is 20K, and the initial learning rate is $10^{-3}$ and decreases twice by factor 10 at iteration 5K and 15K. The iteration number of the second training step is 50K, and the initial learning rate is $10^{-5}$ and decreases by factor 10 at iteration 35K. In addition, the training SNRs are 5 dB and 10 dB, and $\delta$ is set to 20.

For the baseline method, we use QAM which is introduced in Section \ref{sec:2.2} as the constellation mapping step of the full-resolution constellation. The decoder and feature parsing part are trained according to the training strategy.

\begin{figure}[t]

\subfigure{
\begin{minipage}[t]{0.46\linewidth}
  \centering
  \centerline{\includegraphics[width=0.95\linewidth]{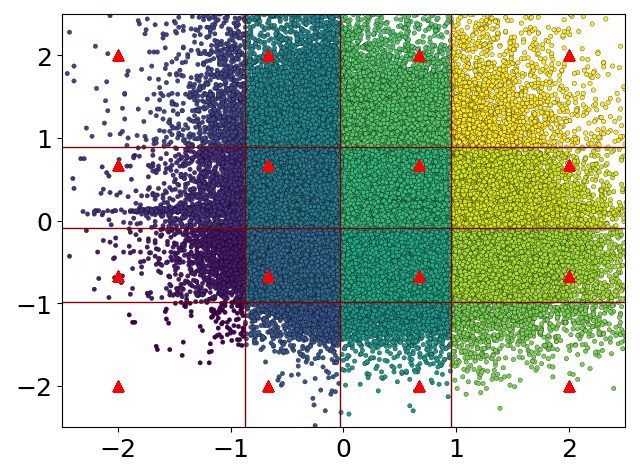}}
  \centerline{(a) Algorithm 1 (MRC)}
\end{minipage}
\label{fig:4(a)}
}
\subfigure{
\begin{minipage}[t]{0.46\linewidth}
  \centering
  \centerline{\includegraphics[width=0.95\linewidth]{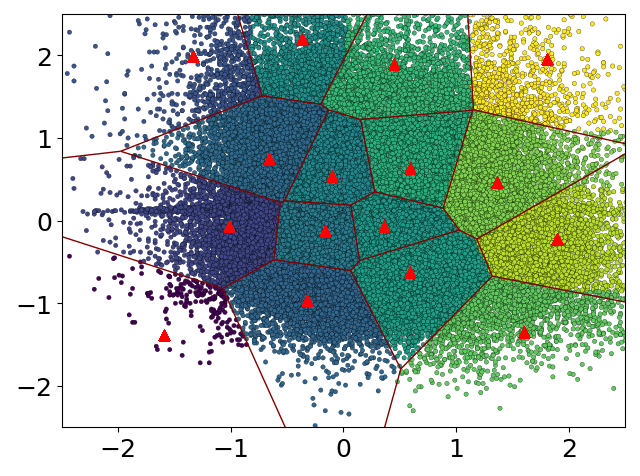}}
  \centerline{(b) Algorithm 2 (MIC)}\medskip
\end{minipage}
\label{fig:4(b)}
}
\caption{The constellation mapping results of the proposed methods.}
\label{fig:4}
\end{figure}

\subsection{Constellation mapping visualization}
\label{sec:5.2}
To show the difference between proposed methods and QAM, we train the models with MRC, MIC, and 16QAM according to the training strategy, and visualize the constellation mapping results in Fig. \ref{fig:4}. The full-resolution constellation is divided into 16 clusters based on different mapping methods. Points in each cluster have the same color and are mapped to a finite constellation point represented by a red triangle.

In 16QAM, all clusters have the same size, and the finite constellation is regular. MRC adjusts the sizes of the clusters based on task performance, but keeps the mapping results the same value as uniform quantization. The cluster sizes in Fig. \ref{fig:4(a)} are different from 16QAM, while the finite constellation is the same as 16QAM. The finite constellation points in MIC are obtained based on task performance, and the clusters are constructed based on clustering. The finite constellation in Fig. \ref{fig:4(b)} is irregularly distributed and the sizes of clusters corresponding to each constellation point are different from the first two methods.

\subsection{Constellation mapping performance}
\label{sec:5.3}
After training the models with different constellation mapping methods under different training SNRs (i.e., SNR$_{train}$), we evaluate the model performance at different testing SNRs (i.e., SNR$_{test}$). As mentioned earlier, our model performs both semantic segmentation and object detection. The quality of predicted segmentation masks is measured with mean intersection over union (mIoU) and the metric for evaluating detection performance is the mean average precision (mAP). 

\begin{figure}[t]
\begin{minipage}[t]{\linewidth}
  \centering
  \centerline{\includegraphics[width=0.95\linewidth]{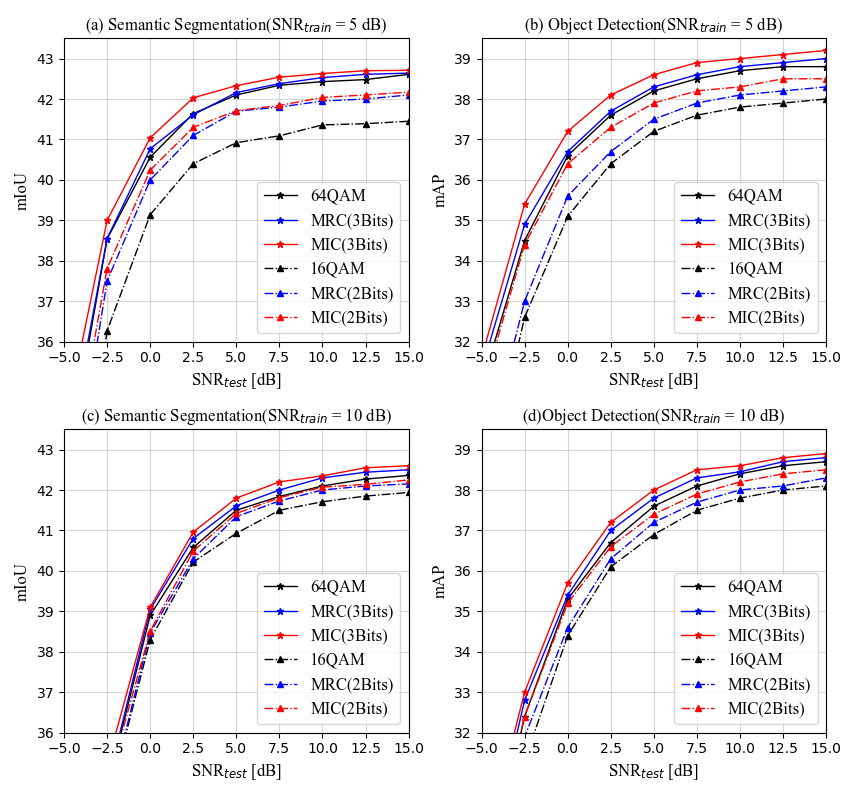}}
\end{minipage}
\caption{The performane of the MTL JSCC with different constellation mapping methods and training SNRs.}
\label{fig:5}
\end{figure}

The performance of the models on two tasks is shown in Fig. \ref{fig:5}. The solid lines are the performance of the proposed methods against 64QAM, and the dashed lines are the performance of the proposed methods against 16QAM. It is clear that both proposed constellation mapping methods outperform QAM on two tasks under different training SNRs. In particular, the performance of MIC is better than that of the other two methods on both tasks under different test SNRs.

Finally, the impact of the algorithms on the computational complexity of the model is analyzed. The computational complexity of the original MTL JSCC model is 127.5 GFLOPs, while the complexity of MRC and MIC is 0.23 MFLOPs and 0.92 MFLOPs, respectively. Therefore, the added module has little effect on the computational complexity of the model.

\section{CONCLUSION}
\label{sec:6}
In this paper, two novel constellation mapping methods are proposed to quantize the full-resolution constellation of the deep JSCC model. Both regular and irregular finite constellations are obtained. With constellation mapping, the deep JSCC model becomes suitable for practical hardware implementation. We apply the constellation mapping methods to the trained MTL JSCC model proposed in our previous work. The performance of the proposed constellation mapping methods on two tasks is better than the traditional QAM mapping. Although the experiments are only carried out on MTL JSCC in this paper, the proposed algorithms can be easily extended to other deep JSCC models.

\balance
\bibliographystyle{IEEEtran}
\bibliography{refs}
\end{document}